\pdfoutput=1
\RequirePackage{fix-cm}
\documentclass[11pt]{article}

\usepackage[T1]{fontenc}
\usepackage[utf8]{inputenc}
\usepackage{mathptmx}
\usepackage{amsmath,amssymb,mathtools}
\usepackage{amsthm}
\usepackage{bm}
\usepackage{booktabs}
\usepackage{array}
\usepackage[margin=1in]{geometry}
\usepackage{graphicx}
\usepackage{tikz}
\usepackage[authoryear]{natbib}
\bibpunct{(}{)}{}{a}{}{;}
\usepackage{hyperref}
\hypersetup{hidelinks}
\usetikzlibrary{arrows.meta,calc,decorations.pathmorphing,patterns,positioning}
\newcommand{\R}{\mathbb R}
\newcommand{\Om}{\Omega}
\newcommand{\dist}{\operatorname{dist}}

\newcommand{\Teff}{\mathcal T}
\newcommand{\Feff}{\mathcal F}
\newtheorem{proposition}{Proposition}

\title{Transition-Set Morphometry for Inter-Scale Contacts in Digital Sandstones}
\author{O. M. Kiselev\\
Innopolis University, Innopolis, Russia\\
\texttt{o.kiselev@innopolis.ru}}

\date{}

\begin{document}

\maketitle

\begin{abstract}
This paper introduces a transition-set morphometry for quantifying inter-scale contacts in digital sandstones. Pore volume, a medial pore-throat subsystem, and fracture porosity are treated as distinct geometric subsystems. Their finite-neighborhood contacts are represented by transition sets such as $T_{VL}$, $T_{VF}$, and, when a residual skeleton is resolved, $T_{LF}$. The method measures the density, finite-scale dimension, and weighted contact measures of these sets. A matrix-mode test was performed on 21 digital sandstone samples from the Imperial College micro-CT collection and Digital Porous Media Portal dataset DRP-317. Permeability, expressed as $\log_{10}K$, served as an external response for assessing how much structural information the descriptors carry. The transition dimension $d_{VL}$ alone produced little independent predictive gain, whereas contact measures combining transition-set density with local pore radius were more informative. The adaptive measure $\widehat C_{VL}^{(d+1)}=\mu_{VL}(\overline D_{VL}/D_*)^{d_{VL}+1}$ increased leave-one-out $R^2$ from 0.752 to 0.881 in the combined sample and from 0.287 to 0.807 within DRP-317. Fractured datasets DRP-5, DRP-31, and DRP-285 were then used to test the transfer of the same transition-set construction to fracture-related contacts. The results support transition sets as reproducible morphometric descriptors of inter-scale contact.

\medskip
\noindent\textbf{Keywords:} digital rock; mathematical morphology; sandstone; transition set; inter-scale contact; finite-scale dimension
\end{abstract}

\section{Introduction}

Porous rocks are commonly described by porosity, pore-size distributions, surface measures, and pore-network parameters. These descriptors emphasize phase volumes and single-subsystem geometry. Inter-subsystem contact requires an explicit finite-scale descriptor, because intergranular pores, pore throats, cemented contacts, microfractures, and larger fractures are generated by different geological processes and occupy different scale ranges. A sample may therefore have a similar porosity and mean pore radius to another sample, but differ in how its pore-throat backbone is embedded in the pore volume or how a fracture is connected to residual matrix porosity.

This work addresses a morphometric question motivated by transport problems in heterogeneous pore systems: can inter-scale contact zones be extracted and measured in a reproducible way from digital rock images? The proposed object is a transition set, defined as the intersection of finite neighborhoods of two previously identified subsystems. For a matrix sandstone, the central transition is $T_{VL}$, which measures the finite-neighborhood contact between the pore volume $V$ and a medial pore-throat subsystem $L$. For a fractured volume, analogous transitions include $T_{VF}$, between residual porosity and an observed fracture component, and $T_{FM}$, between a thresholded fracture and the surrounding matrix.

The construction is related to several established lines of work. Fractal and multifractal analyses have long been used to describe pore structures, fractures, nuclear magnetic resonance data, and mercury-intrusion porosimetry in geological media \citep{Mandelbrot1982,Falconer2003,HentschelProcaccia1983,Halsey1986,Bonnet2001,Giri2015,QuGuo2020,Lai2019}. Mathematical morphology provides the language of neighborhoods, dilations, erosions, skeletons, and intersections \citep{Matheron1975,Serra1982}. Digital rock physics uses these operations in image analysis, skeletonization, and pore-network extraction \citep{OrenBakke2002,ValvatneBlunt2004,DongBlunt2009,Blunt2013,Gostick2019}. Minkowski functionals and related integral-geometric descriptors measure the volume, surface, curvature, and connectivity of a selected phase, while ordinary box counting measures the dimension of a single set. The descriptor used here is a finite-scale contact zone between two subsystems.

The contribution is threefold. First, a general transition-set formalism is introduced for finite-scale contacts between pore, throat, node, matrix, and fracture subsystems. Second, a reproducible workflow is implemented for matrix sandstones, and the resulting transition descriptors are compared with porosity, mean pore radius, and one-phase dimensions under leave-one-out validation. Third, observed fractured datasets are used to show how the same construction transfers from a matrix pore-throat contact to fracture-matrix and fracture-residual-porosity contacts. Together, these steps define a compact morphometric layer for digital-rock characterization.

The paper is organized as follows. Sect.~\ref{sec:geometry} introduces the scale hierarchy and transition sets. Sect.~\ref{sec:measures} defines density, finite-scale dimension, generalized dimensions, and weighted contact measures. Sect.~\ref{sec:matrix_fracture} specifies the matrix and fracture transitions used in the computations. Sect.~\ref{sec:matrix_results} presents the matrix-mode validation against $\log_{10}K$. Sect.~\ref{sec:fracture_results} applies the same construction to fractured digital rock datasets. Sect.~\ref{sec:conclusions} summarizes the implications and limitations.

\section{Geometric Setting}
\label{sec:geometry}

Let a sandstone sample be represented by a bounded digital domain $\Om\subset\R^3$ containing solid grains, pore space, cemented contacts, and, when present, fractures. The geometry is described by a hierarchy of characteristic scales
\[
    L_0\gg L_1\gg\cdots\gg L_N,
\]
where $L_0$ corresponds to macroscopic reservoir heterogeneity and $L_N$ to small pores, pore throats, microfractures, and internal surface roughness. A typical sandstone hierarchy includes fault and damage-zone scales, fracture networks, grain contacts, intergranular pores, pore throats, and submicron cement or clay porosity. The numerical boundaries between these levels provide an order-of-magnitude scale frame in which the geometric type of a defect or pore element changes.

At a scale level $L_n$, let $\Feff_n\subset\Om$ denote a finite-scale geometric set. It may be decomposed into components
\[
    \Feff_n=
    \Feff_n^{(V)}\cup
    \Feff_n^{(S)}\cup
    \Feff_n^{(L)}\cup
    \Feff_n^{(X)} ,
\]
where $V$ denotes volumetric pore clusters, $S$ internal surfaces or crack-like sheets, $L$ pore throats and channel-like elements, and $X$ nodes, endpoints, and singular contacts. Their dimensions satisfy the qualitative ranges
\[
    2<d_n^{(V)}\le 3,\qquad
    2\le d_n^{(S)}<3,\qquad
    1\le d_n^{(L)}<2,\qquad
    0\le d_n^{(X)}<1.
\]
These ranges serve as qualitative guides and clarify why a single effective dimension cannot represent all geometrically relevant elements of a sandstone.

For a component $\Feff_n^{(a)}$, a local box-counting relation in a window $B_L(x)$ is written as
\[
    N_n^{(a)}(\ell;x,L)
    \sim
    \left(\frac{L}{\ell}\right)^{d_n^{(a)}(x,L)} ,
    \qquad a\in\{V,S,L,X\}.
\]
Here $N_n^{(a)}(\ell;x,L)$ is the number of cells of size $\ell$ that intersect the component in the window. A nearly linear object gives $d\approx1$, a surface-like object gives $d\approx2$, and a volumetric pore cluster gives $d$ close to 3. Fractional values indicate heterogeneous finite-scale occupation.

To describe the distribution of measure across occupied cells, let $\nu_n^{(a)}$ be a measure assigned to $\Feff_n^{(a)}$ and define
\[
    m_{n,i}^{(a)}(\ell)
    =
    \frac{
        \nu_n^{(a)}\bigl(\Feff_n^{(a)}\cap Q_i(\ell)\bigr)
    }{
        \nu_n^{(a)}\bigl(\Feff_n^{(a)}\bigr)
    } .
\]
The values $m_{n,i}^{(a)}$ are cell-wise fractions of the selected measure and satisfy $\sum_i m_{n,i}^{(a)}=1$. Their moment sum is
\[
    Z_n^{(a)}(q,\ell)
    =
    \sum_i
    \left(m_{n,i}^{(a)}(\ell)\right)^q
    \sim
    \ell^{\tau_n^{(a)}(q)} .
\]
The generalized dimension is
\[
    D_{q,n}^{(a)}=\frac{\tau_n^{(a)}(q)}{q-1},
    \qquad q\ne 1,
\]
with the information dimension obtained by the usual limiting expression
\[
    D_{1,n}^{(a)}
    =
    \lim_{\ell\to0}
    \frac{
        \sum_i m_{n,i}^{(a)}(\ell)\log m_{n,i}^{(a)}(\ell)
    }{
        \log \ell
    } .
\]
In a digital image, all these quantities are finite-scale slopes over a prescribed set of cell sizes.

For two components $\Feff_n^{(a)}$ and $\Feff_m^{(b)}$, the $\delta$-neighborhood is
\[
    (\Feff_n^{(a)})_\delta
    =
    \{x\in\Om:\dist(x,\Feff_n^{(a)})<\delta\}.
\]
The transition set is then
\[
    \Teff_{nm}^{(ab)}(\delta)
    =
    (\Feff_n^{(a)})_\delta
    \cap
    (\Feff_m^{(b)})_\delta .
\]
This set records the finite-thickness region where the two subsystems are mutually close at scale $\delta$. Repeating the construction over several $\delta$ values gives a scale-growth curve for the contact.

\begin{proposition}[Transition hierarchy]
The geometry of a digital sandstone can be represented by finite-scale sets $\{\Feff_n,d_n\}$ and by transition sets $\{\Teff_{nm}^{(ab)},d_{nm}^{(ab)}\}$. The transition sets measure where subsystems of different geometric type or scale are in finite-neighborhood contact and allow node-like, line-like, surface-like, and volume-distributed contacts to be separated.
\end{proposition}

\section{Transition-Set Measures}
\label{sec:measures}

For a region of interest $Q\subset\Om$, the density of a transition set is
\[
    \mu_{nm}(Q;\delta)=\frac{|T_{nm}(\delta)\cap Q|}{|Q|}.
\]
The denominator is the full digital volume of the region of interest, so that densities can be compared across samples. For the matrix pore-throat transition used below,
\[
    \mu_{VL}(Q)
    =
    \frac{|(L_m)_{\delta_L}\cap V_m\cap Q|}{|Q|},
\]
which is the fraction of $Q$ occupied both by pore volume and by the finite neighborhood of the medial pore-throat subsystem. For a fractured binary volume,
\[
    \mu_{VF}(Q)
    =
    \frac{|(V_{\mathrm{res}})_\delta\cap(F_{\mathrm{obs}})_\delta\cap Q|}{|Q|}.
\]
The finite-scale dimension $d_{nm}$ is computed by box counting on the same transition set and describes whether the contact behaves more like a node, a curve, a surface, or a volume-distributed region over the selected scale range.

A density alone omits local size information. A weighted contact measure is therefore defined as
\[
    C_{nm}^{[P]}(\delta)
    =
    \frac{1}{|\Omega|}
    \sum_{x\in T_{nm}(\delta)} w_P(x),
\]
where $P$ denotes the chosen diagnostic scenario and $w_P(x)$ is a local weight. In continuous notation, this expression corresponds to an integral over $T_{nm}(\delta)$. If $w_P(x)\equiv1$, the quantity reduces to the transition density. If the local pore radius, fracture aperture, or another length scale is relevant, it can be incorporated through
\[
    w_P(x)
    =
    \chi_P(x)
    \left(
        \frac{\lambda_P(x)}{\lambda_*}
    \right)^{\beta_P},
\]
where $\lambda_P(x)$ is a local length, $\lambda_*$ is a reference length, $\beta_P$ is a specified exponent, and $\chi_P(x)$ is an optional activity indicator.

For the matrix transition $T_{VL}$, the Euclidean distance-transform radius
\[
    r(x)=\dist(x,\Om\setminus V)
\]
is used as the local pore scale. The radius-weighted measures are
\[
    C_{VL}^{(p)}
    =
    \frac{1}{|\Omega_{\mathrm{ROI}}|}
    \sum_{x\in T_{VL}} r(x)^p,
    \qquad p=2,4.
\]
The case $p=2$ gives an area-like weighting, while $p=4$ is a Poiseuille-type diagnostic weighting. Their hydraulic interpretation depends on the flow model supplied with the morphometric descriptor.

An aggregated dimension-weighted form is also used:
\[
    \widehat C_{nm}^{[P]}(\delta)
    =
    \mu_{nm}(\Omega;\delta)
    \left(
        \frac{\overline\lambda_{nm}^{[P]}(\delta)}{\lambda_*}
    \right)^{\alpha_{nm}^{[P]}} .
\]
For the matrix pore-throat transition,
\[
    \overline D_{VL}=2\,\langle r(x)\rangle_{x\in T_{VL}},
\]
and
\[
    \widehat C_{VL}^{(d+1)}
    =
    \mu_{VL}
    \left(
        \frac{\overline D_{VL}}{D_*}
    \right)^{d_{VL}+1}.
\]
In the calculations, $D_*=1\,\mu\mathrm m$. The exponent $d_{VL}+1$ is introduced as a geometric and dimensional heuristic. The motivation is that a transition set with dimension $d_{VL}$ has an approximate cross-sectional scaling of $d_{VL}-1$, and a laminar mobility factor contributes a length-squared factor in canonical geometries. This gives $D^{(d_{VL}-1)+2}=D^{d_{VL}+1}$. The limiting cases are compatible with slit-like $D^3$ and tube-like $D^4$ scaling. The numerical test below evaluates this feature as a morphometric diagnostic.

\section{Matrix and Fracture Transitions}
\label{sec:matrix_fracture}

For a binary micro-CT image of a matrix sandstone, the pore volume is denoted by $V_m$, the internal pore surface by $S$, the medial pore-throat subsystem by $L_m$, and the nodes of that subsystem by $X_m$. The baseline one-phase descriptors are porosity $\phi$ and finite-scale dimensions $d_V$, $d_S$, and $d_L$. The proposed additional descriptors are based on
\[
    T_{VL}=(L_m)_{\delta_L}\cap V_m,
    \qquad
    T_{LX}=(X_m)_{\delta_X}\cap (L_m)_{\delta_L}.
\]
In $T_{VL}$, the undilated pore volume restricts the neighborhood of the medial subsystem to the pore phase and thus preserves the operational meaning of a pore-to-medial contact.

For an observed fractured binary volume, let $V_{\mathrm{void}}$ be the full void phase. The largest connected void component is denoted by $F_{\mathrm{obs}}$ and interpreted operationally as the observed fracture component when a residual void phase remains. The residual porosity is
\[
    V_{\mathrm{res}}=V_{\mathrm{void}}\setminus F_{\mathrm{obs}}.
\]
The primary fracture transition is
\[
    T_{VF}=(V_{\mathrm{res}})_\delta\cap(F_{\mathrm{obs}})_\delta.
\]
If a stable residual pore-throat skeleton can be extracted, a finer transition
\[
    T_{LF}=(L_{\mathrm{res}})_\delta\cap(F_{\mathrm{obs}})_\delta
\]
can be used. In the available fractured data, residual-skeleton extraction is unstable, so $T_{VF}$ remains the main measured object. For the full-core grayscale control dataset DRP-285, matrix microporosity lies below the scan resolution, and the measured transition is
\[
    T_{FM}=(F)_\delta\cap(M)_\delta,
\]
where $F$ is a thresholded fracture phase and $M$ is its complement inside the region of interest.

The matrix-mode comparison uses $\log_{10}K$ as an external response. Permeability values are taken from existing computational or laboratory metadata and are used to test whether the transition descriptors carry information beyond porosity and mean pore radius. The logarithm is used because sandstone permeability often varies by orders of magnitude, and a power-law relation between a geometric feature and permeability becomes approximately linear after logarithmic transformation.

\section{Matrix-Mode Computation and Results}
\label{sec:matrix_results}

The matrix-mode analysis used two sandstone sources. The first consists of Berea and samples S1--S9 from the Imperial College Micro-CT Images and Networks collection \citep{ImperialImagesNetworks}. The second is DRP-317 from the Digital Porous Media Portal, which contains 11 Kocurek sandstones with segmented $1000^3$ voxel images at $2.25\,\mu\mathrm m$ resolution and laboratory gas permeability values \citep{DRP317,Neumann2021}. Table~\ref{tab:unified_sources_en} summarizes the sources.

\begin{table}[htbp]
\centering
\small
\setlength{\tabcolsep}{3pt}
\begin{tabular}{@{}p{0.25\linewidth}ccp{0.23\linewidth}p{0.17\linewidth}@{}}
\toprule
Source & $N$ & Size & Target response & Processing \\
\midrule
Imperial Micro-CT Images and Networks & $10$ & $300^3$--$400^3$ & image-based $K_{\mathrm{avg}}$ from metadata & full volume \\
DRP-317 Kocurek sandstones & $11$ & $1000^3$ & laboratory gas permeability & central $512^3$ ROI \\
\bottomrule
\end{tabular}
\caption{Independent sandstone sources used for matrix-mode morphometry}
\label{tab:unified_sources_en}
\end{table}

For each binary image, the largest connected pore component was retained. The internal surface $S$ was computed as the boundary of $V$ relative to a six-connected erosion. The distance transform $r(x)=\dist(x,\Om\setminus V)$ was used to define a medial ridge subsystem $L$: a pore voxel was assigned to $L$ if $r(x)$ was a local maximum in several opposite directions in the 26-neighborhood and $r(x)\ge1.5$ voxels. The node set $X$ consisted of ridge voxels with at least five ridge neighbors. The transition radii were $\delta_L=2$ voxels and $\delta_X=3$ voxels.

Box-counting dimensions were computed on common physical scales
\[
    18,\ 36,\ 72,\ 144,\ 288\,\mu\mathrm m.
\]
For each set, the number of occupied cells $N(s)$ was fitted by
\[
    \log N(s)=d\,\log\frac{n_{\mathrm{ref}}}{s}+b,
    \qquad
    n_{\mathrm{ref}}=\min(n_x,n_y,n_z).
\]
Generalized dimensions were computed for $q=0,1,2$.

Linear models were fitted to $\log_{10}K$. For the combined sample, the baseline model included source and porosity,
\[
    \log_{10}K
    =
    a_0+a_s I_{\mathrm{DRP}}+a_\phi\phi+\varepsilon.
\]
Candidate features were then added one at a time. The quality metric was leave-one-out $R^2$, denoted by $R^2_{-1}$:
\[
    R^2_{-1}
    =
    1-\frac{\sum_i(y_i-\widehat y_{-i})^2}{\sum_i(y_i-\overline y)^2},
    \qquad
    y_i=\log_{10}K_i.
\]
For added features, a permutation control was performed by randomly permuting $\log_{10}K$ among samples 4999 times and recomputing the gain in $R^2_{-1}$. The value $p_\pi$ is the right-tail fraction of permutations whose gain was at least as large as the observed gain.

\begin{table}[htbp]
\centering
\small
\begin{tabular}{p{0.33\linewidth}cccc}
\toprule
Model & $N=21$ & $p_\pi$ & Imperial & DRP-317 \\
 & $R^2_{-1}$ &  & $R^2_{-1}$ & $R^2_{-1}$ \\
\midrule
source, $\phi$ & $0.752$ & -- & $0.283$ & $0.287$ \\
source, $\phi,d_V$ & $0.790$ & $0.139$ & $-0.077$ & $0.301$ \\
source, $\phi,d_S$ & $0.825$ & $0.099$ & $0.173$ & $0.334$ \\
source, $\phi,d_L$ & $0.715$ & $0.256$ & $0.257$ & $0.228$ \\
source, $\phi,\mu_L$ & $0.779$ & $0.145$ & $0.083$ & $0.116$ \\
source, $\phi,d_{VL}$ & $0.757$ & $0.181$ & $-0.069$ & $0.261$ \\
source, $\phi,\mu_{VL}$ & $0.809$ & $0.117$ & $0.264$ & $0.174$ \\
source, $\phi,\log C_{VL}^{(2)}$ & $0.894$ & $0.066$ & $0.619$ & $0.785$ \\
source, $\phi,\log C_{VL}^{(4)}$ & $0.909$ & $0.053$ & $0.664$ & $0.697$ \\
source, $\phi,\log\widehat C_{VL}^{(d+1)}$ & $0.881$ & $0.073$ & $0.519$ & $0.807$ \\
source, $\phi,d_{VL},D_{1,VL}$ & $0.732$ & -- & $-0.386$ & $0.540$ \\
source, $\phi,\bar r$ & $0.836$ & $0.102$ & $0.522$ & $0.659$ \\
\bottomrule
\end{tabular}
\caption{Leave-one-out comparison with $\log_{10}K$. Combined models include a source indicator; $p_\pi$ is the permutation level for the gain over the source-plus-porosity baseline}
\label{tab:unified_regression_en}
\end{table}

The transition dimension $d_{VL}$ alone adds little information to porosity and source. In the combined sample, the gain is $\Delta R^2_{-1}=0.004$ with $p_\pi=0.181$, and within DRP-317 the model with $d_{VL}$ gives $R^2_{-1}=0.261$, below the porosity-only value of 0.287. One-phase dimensions show heterogeneous behavior. For example, $d_S$ gives $R^2_{-1}=0.825$ in the combined sample, with a permutation level of 0.099 and source-dependent performance.

The main matrix-mode signal appears when the amount of transition is combined with local pore-size information. The measures $\log C_{VL}^{(2)}$, $\log C_{VL}^{(4)}$, and $\log\widehat C_{VL}^{(d+1)}$ increase the combined-sample $R^2_{-1}$ to 0.894, 0.909, and 0.881, respectively. The corresponding permutation levels are 0.066, 0.053, and 0.073. These values indicate diagnostic predictive evidence in the combined heterogeneous sample. Within DRP-317, the same measures give stronger gains relative to porosity alone, with $R^2_{-1}=0.785$, 0.697, and 0.807.

The mean pore radius is a particularly important baseline. In the combined sample, adding $\bar r$ gives $R^2_{-1}=0.836$, so the gain from $\bar r$ to $\widehat C_{VL}^{(d+1)}$ is modest, $0.836\to0.881$, and the gain to $C_{VL}^{(4)}$ is $0.836\to0.909$. The result indicates that the transition zone contains additional interface information beyond the mean pore scale.

\begin{table}[htbp]
\centering
\footnotesize
\begin{tabular}{@{}c c c c c@{}}
\toprule
$r_{\min}$, vox. &
$\langle d_{VL}+1\rangle$ &
$N=21$, $R^2_{-1}$ &
DRP-317, $R^2_{-1}$ &
$N=21$, $R^2_{-1}$ for $C_{VL}^{(4)}$ \\
& range over $\delta_L$ &
for $\widehat C_{VL}^{(d+1)}$ &
for $\widehat C_{VL}^{(d+1)}$ &
range over $\delta_L$ \\
\midrule
$1.0$ & $3.715$--$3.716$ & $0.824$--$0.872$ & $0.795$--$0.818$ & $0.905$--$0.909$ \\
$1.5$ & $3.656$--$3.707$ & $0.881$--$0.889$ & $0.793$--$0.826$ & $0.906$--$0.909$ \\
$2.0$ & $3.635$--$3.701$ & $0.887$--$0.904$ & $0.793$--$0.829$ & $0.907$--$0.910$ \\
$2.5$ & $3.514$--$3.636$ & $0.927$--$0.934$ & $0.800$--$0.841$ & $0.912$--$0.913$ \\
\bottomrule
\end{tabular}
\caption{Sensitivity of weighted transition measures to the medial-radius threshold $r_{\min}$ and transition-neighborhood radius $\delta_L$}
\label{tab:skeleton_sensitivity_en}
\end{table}

Table~\ref{tab:skeleton_sensitivity_en} shows that the diagnostic gain of $\widehat C_{VL}^{(d+1)}$ is retained over the tested parameter grid. However, the absolute values depend on $r_{\min}$ and $\delta_L$. The medial ridge subsystem is sensitive to resolution, segmentation noise, and the rule used to define local maxima of the distance field. Therefore, $L_m$, $T_{VL}$, and their derived descriptors are workflow-dependent morphometric quantities and are reported together with parameter sensitivity.

\section{Fracture-Mode Transfer}
\label{sec:fracture_results}

The 21 matrix sandstones were first screened for possible fracture-like pore blocks. Each region of interest was divided into blocks of physical size about $72\,\mu\mathrm m$. In pore-containing blocks, the shape tensor eigenvalues $\lambda_1\ge\lambda_2\ge\lambda_3$ were used to compute
\[
    P=\frac{\lambda_2-\lambda_3}{\lambda_1},\qquad
    L=\frac{\lambda_1-\lambda_2}{\lambda_1},\qquad
    S=\frac{\lambda_3}{\lambda_1}.
\]
Blocks were marked as fracture-like candidates when
\[
    P\ge0.45,\qquad L\le0.55,\qquad S\le0.12,
    \qquad \widetilde r\le12\,\mu\mathrm m,
\]
where $\widetilde r$ is the median pore radius in the block.

\begin{table}[htbp]
\centering
\footnotesize
\begin{tabular}{@{}p{0.23\linewidth}p{0.23\linewidth}p{0.24\linewidth}p{0.18\linewidth}@{}}
\toprule
Sample & Candidate blocks among pore blocks & Candidate porosity in ROI & Largest domain, $\mu\mathrm m$ \\
\midrule
S7 & $7.59\%$ & $0.92\%$ & $504$ \\
S5 & $7.43\%$ & $0.68\%$ & $432$ \\
Leopard & $6.82\%$ & $0.72\%$ & $504$ \\
S2 & $6.31\%$ & $0.79\%$ & $297$ \\
Bentheimer & $5.86\%$ & $0.72\%$ & $360$ \\
\bottomrule
\end{tabular}
\caption{Most pronounced fracture-like candidates from the block screening}
\label{tab:microcrack_screening_en}
\end{table}

The candidate porosity remains small, ranging from $0.07$ to $0.92\%$ of the region of interest over the full sample set. Additional connected-component analysis inside candidate blocks indicates that the largest components are often linear or mixed, with plate-like components contributing less prominently. The 21 samples are therefore treated as a matrix background in the subsequent analysis.

The fracture-mode test uses observed fractured datasets DRP-5 and DRP-31 as independent objects \citep{DRP5,TokanLawal2015,DRP31,Karpyn2007}. Their segmentations are binary, so the separation into fracture and residual porosity is operational. The largest connected component of $V_{\mathrm{void}}$ is treated as $F_{\mathrm{obs}}$, and the remaining void space is $V_{\mathrm{res}}$. This rule is a morphological segmentation criterion.

\begin{table}[htbp]
\centering
\scriptsize
\setlength{\tabcolsep}{2pt}
\begin{tabular}{@{}p{0.15\linewidth}ccccccp{0.14\linewidth}@{}}
\toprule
Volume & $\phi_{\mathrm{void}}$ & $\mu_F$ & $\mu_{V,\mathrm{res}}$ & $V_{\mathrm{res}}/V_{\mathrm{void}}$ & $d_F$ & $d_{V,\mathrm{res}}$ & Interpretation \\
\midrule
DRP-5 TP1 & $0.0676$ & $0.0418$ & $0.0258$ & $0.382$ & $2.86$ & $2.60$ & fracture and residual porosity \\
DRP-5 TP2 & $0.1073$ & $0.0854$ & $0.0219$ & $0.204$ & $2.89$ & $2.52$ & fracture and residual porosity \\
DRP-31 Berea & $0.0945$ & $0.0945$ & $0$ & $0$ & $2.70$ & -- & one connected void component \\
\bottomrule
\end{tabular}
\caption{Operational decomposition of void space in observed fractured volumes}
\label{tab:real_fractured_decomposition_en}
\end{table}

Table~\ref{tab:real_fractured_decomposition_en} shows that DRP-5 contains a separable pair consisting of a large fracture-like component and residual void space. For TP1 the largest component accounts for 61.8\% of void space, and for TP2 it accounts for 79.6\%. DRP-31 is a limiting case in which the accepted rule produces one connected void component and no separable residual matrix phase.

For DRP-5, the measured transition is
\[
    T_{VF}=(V_{\mathrm{res}})_\delta\cap(F_{\mathrm{obs}})_\delta.
\]
At $\delta=2$ voxels, the transition density and dimension are
\[
    \mu_{VF}=0.026,\quad d_{VF}=2.63
    \qquad \text{for TP1},
\]
and
\[
    \mu_{VF}=0.018,\quad d_{VF}=2.32
    \qquad \text{for TP2}.
\]
The fraction of the fracture component lying in the $\delta$-neighborhood of residual porosity is 5.9\% for TP1 and 2.4\% for TP2. The fraction of residual porosity lying near the fracture is similar in the two subsets, 7.4\% and 7.5\%, respectively.

\begin{figure}[htbp]
\centering
\includegraphics[width=\linewidth]{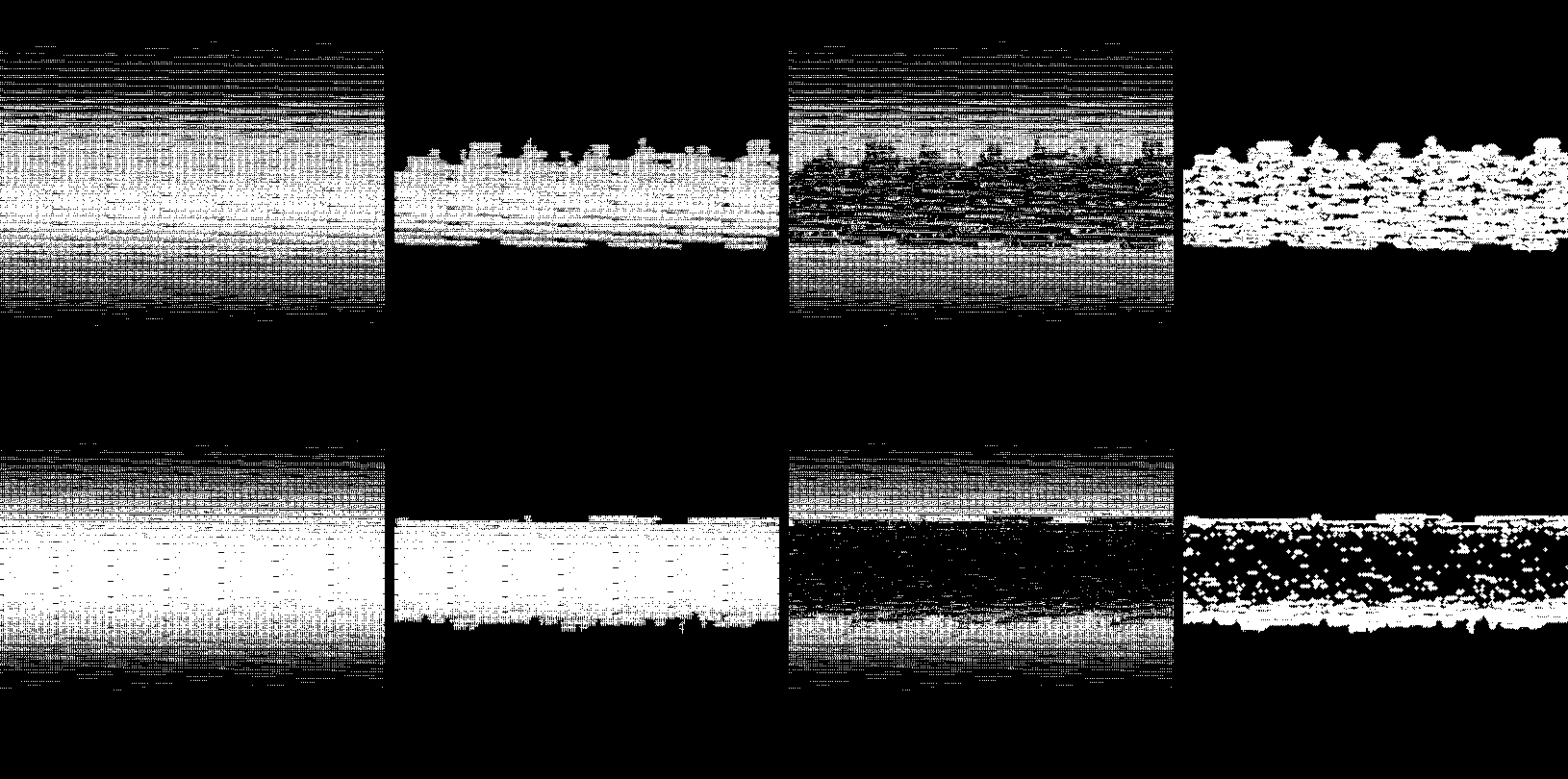}
\caption{Binary projections of DRP-5 TP1 and TP2 showing full void space, the largest fracture-like component, residual porosity, and $T_{VF}$ at $\delta=2$ voxels}
\label{fig:drp5_real_transition_en}
\end{figure}

\begin{figure}[htbp]
\centering
\includegraphics[width=0.92\linewidth]{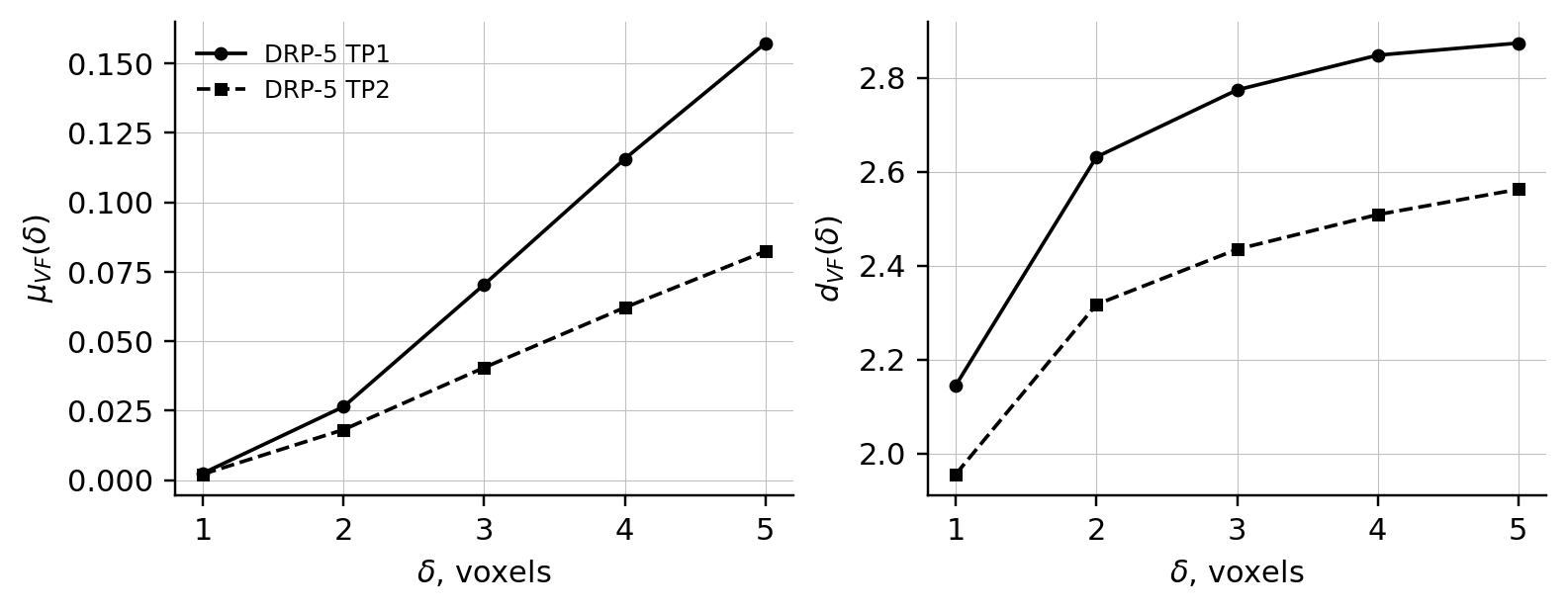}
\caption{Scale-growth curve of the transition set $T_{VF}$ in DRP-5}
\label{fig:drp5_vf_delta_growth_en}
\end{figure}

\begin{table}[htbp]
\centering
\footnotesize
\begin{tabular}{@{}c cc cc@{}}
\toprule
$\delta$, vox. &
\multicolumn{2}{c}{DRP-5 TP1} &
\multicolumn{2}{c}{DRP-5 TP2} \\
\cmidrule(lr){2-3}\cmidrule(l){4-5}
 & $\mu_{VF}$ & $d_{VF}$ & $\mu_{VF}$ & $d_{VF}$ \\
\midrule
$1$ & $0.0025$ & $2.15$ & $0.0022$ & $1.96$ \\
$2$ & $0.0264$ & $2.63$ & $0.0182$ & $2.32$ \\
$3$ & $0.0704$ & $2.77$ & $0.0405$ & $2.43$ \\
$4$ & $0.1156$ & $2.85$ & $0.0621$ & $2.51$ \\
$5$ & $0.1572$ & $2.87$ & $0.0823$ & $2.56$ \\
\bottomrule
\end{tabular}
\caption{Scale curve of the measured transition $T_{VF}$ in DRP-5}
\label{tab:real_vf_delta_en}
\end{table}

The fracture-contact exponent $\alpha=d+1$ falls between the slit-like and pore-like scaling ranges. At $\delta=2$ voxels, $\alpha_{VF}=3.63$ for TP1 and 3.32 for TP2. These values serve as morphological controls on the adaptive exponent. Residual-skeleton extraction in DRP-5 at $r\ge1.5$ voxels is unstable; lowering the threshold to $r=1$ makes the ridge subsystem nearly coincide with all residual porosity. Thus $T_{VF}$ is the accessible inter-scale object in these data.

Dataset DRP-285 provides a full-core grayscale control for a tight Chang 6 sandstone fractured by different fluids \citep{DRP285,Yang2021Fracture}. The associated experiment used cylindrical samples with diameter $50\,\mathrm{mm}$ and height $100\,\mathrm{mm}$, a central injection hole of $9\,\mathrm{mm}$, and micro-CT at $53\,\mu\mathrm m$ resolution \citep{Yang2021Fracture}. A central $128^3$ voxel region of interest was selected from the post-fracturing water stack, aligned with the visible fracture. Otsu thresholding inside the region gave $\tau_O=13876$. The low-intensity phase below this threshold was treated as fracture porosity $F$, and the complement as matrix $M$.

\begin{table}[htbp]
\centering
\scriptsize
\setlength{\tabcolsep}{2pt}
\begin{tabular}{@{}p{0.22\linewidth}p{0.18\linewidth}cccc@{}}
\toprule
Stack & State & TIFF slices & Slice size & Format & Size, MiB \\
\midrule
Original DRP-285 volume & before loading & $1341$ & $1001\times1024$ & float32 & $5244.0$ \\
Water-fractured DRP-285 & after water fracturing & $937$ & $999\times1024$ & float32 & $3656.8$ \\
\bottomrule
\end{tabular}
\caption{Inventory of available DRP-285 stacks used for the full-core grayscale control}
\label{tab:drp285_inventory_en}
\end{table}

At $\tau_O$, the fracture phase is connected and occupies 7.38\% of the region of interest. Its finite-scale dimension is $d_F=2.34$, and the shape indicators are $P=0.535$, $L=0.288$, and $S=0.178$. Threshold sensitivity over $0.9\tau_O$ and $1.1\tau_O$ gives $\mu_F=5.55$--9.65\% and $d_F=2.28$--2.41. The measured transition is
\[
    T_{FM}=(F)_\delta\cap(M)_\delta.
\]

\begin{table}[htbp]
\centering
\footnotesize
\begin{tabular}{@{}c cccc@{}}
\toprule
$\delta$, vox. & $\delta$, $\mu\mathrm m$ & $\mu_{FM}$ & $d_{FM}$ & Fraction of $F$ near $M$ \\
\midrule
$1$ & $53$ & $0.0470$ & $2.33$ & $0.317$ \\
$2$ & $106$ & $0.0913$ & $2.40$ & $0.602$ \\
$3$ & $159$ & $0.1296$ & $2.46$ & $0.811$ \\
$4$ & $212$ & $0.1608$ & $2.51$ & $0.927$ \\
$5$ & $265$ & $0.1859$ & $2.54$ & $0.964$ \\
\bottomrule
\end{tabular}
\caption{Scale curve of the fracture-matrix transition $T_{FM}$ in the DRP-285 region of interest}
\label{tab:drp285_fm_delta_en}
\end{table}

\begin{figure}[htbp]
\centering
\includegraphics[width=0.92\linewidth]{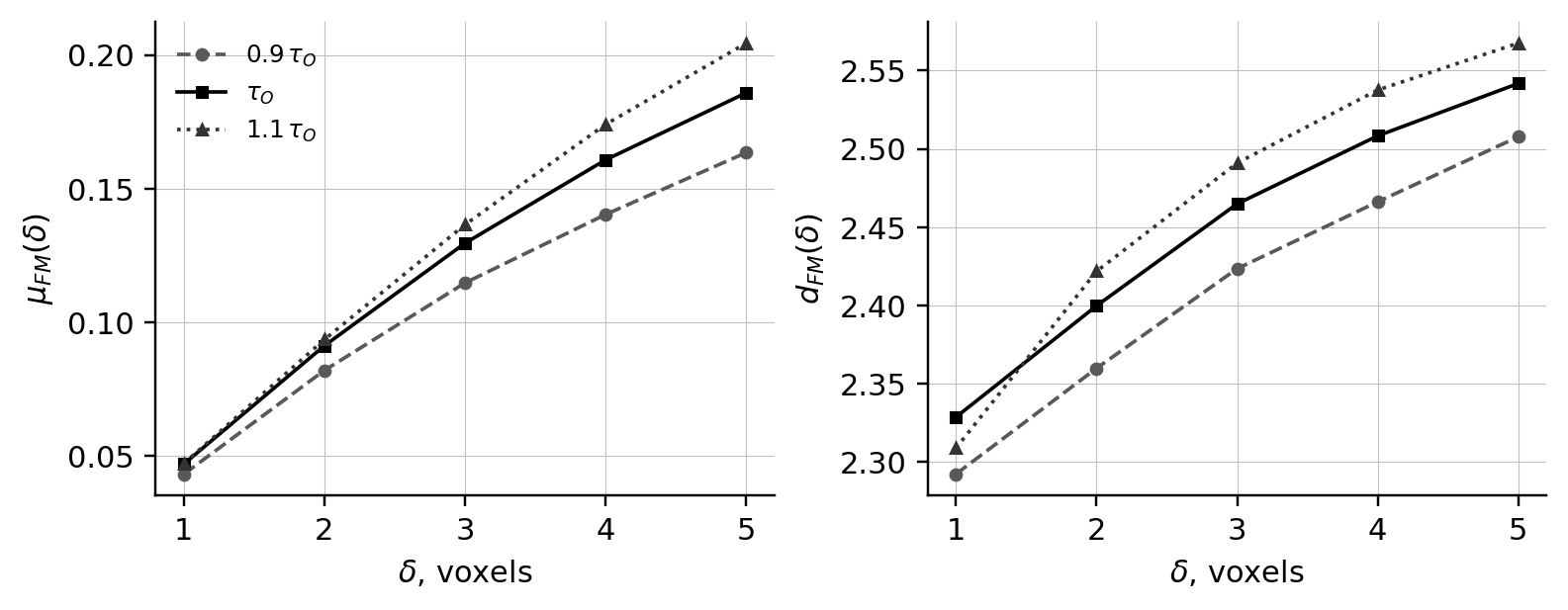}
\caption{Scale-growth curve of $T_{FM}$ in the DRP-285 fracture region; dashed curves show threshold sensitivity}
\label{fig:drp285_fm_delta_growth_en}
\end{figure}

For DRP-285, $d_{FM}+1=3.40$ at $\delta=2$ voxels and 3.46 at $\delta=3$ voxels. This supports the interpretation that $T_{FM}$ describes a macrofracture-matrix contact at the full-core grayscale resolution.

\section{Discussion}
\label{sec:discussion}

The matrix and fracture analyses use the same finite-neighborhood construction in two geometric settings. In the matrix mode, transition descriptors are compared with an external permeability response. In the fracture mode, the same operation is applied to fracture-related contacts. The common element is the construction of a finite-neighborhood contact set between two subsystems. In the matrix case this is $T_{VL}$, a contact between pore volume and a medial pore-throat subsystem. In the fracture case it is $T_{VF}$ or $T_{FM}$, a contact between a fracture component and residual porosity or matrix.

The strongest limitation is the small and heterogeneous validation set. The combined matrix sample has 21 images from two sources, and the permeability targets were obtained by different procedures. The combined-sample regression results are therefore external diagnostic comparisons. The bootstrap intervals are wide, especially within individual sources. The fracture datasets provide a morphological validation of contact extraction.

Skeleton dependence is another central limitation. The medial ridge subsystem is sensitive to voxel resolution, segmentation noise, and the chosen distance-transform ridge rule. The sensitivity analysis shows that the main weighted transition measures remain above the corresponding baselines over the tested parameter grid, but the absolute $R^2_{-1}$ values change with $r_{\min}$ and $\delta_L$. The reported descriptors are therefore tied to the image-processing definitions and parameter sensitivity used in the analysis.

The main practical implication is that transition-set morphometry extends standard digital-rock descriptors. Porosity and mean pore radius remain strong and interpretable baseline features. The transition measures add an interface layer: they quantify where a medial pore-throat subsystem, fracture component, or matrix phase enters into finite-scale contact with another subsystem.

\section{Conclusions}
\label{sec:conclusions}

A transition-set morphometry has been introduced for inter-scale contacts in digital sandstones. The basic construction
\[
    \Teff_{nm}(\delta)=(\Feff_n)_\delta\cap(\Feff_m)_\delta
\]
extracts finite-neighborhood contacts between predefined geometric subsystems. These contacts are described by density, finite-scale dimension, generalized dimensions, and optional process-weighted measures.

In 21 matrix sandstone images, the transition dimension $d_{VL}$ by itself gave a weak independent signal relative to porosity. The more informative descriptors were measures that combine the amount of pore-to-medial contact with local pore size. The adaptive measure $\widehat C_{VL}^{(d+1)}$ increased leave-one-out $R^2$ from 0.752 to 0.881 in the combined sample and from 0.287 to 0.807 within DRP-317. The radius-weighted measure $C_{VL}^{(4)}$ gave a combined-sample value of 0.909. These results are reported as diagnostic morphometric evidence.

Fractured datasets showed that the same operation transfers to fracture-related contacts. In DRP-5, the transition $T_{VF}$ quantified contact between the largest fracture-like void component and residual porosity. In DRP-285, the transition $T_{FM}$ quantified contact between a thresholded macrofracture and surrounding matrix at full-core scale. These analyses demonstrate morphological transferability across matrix and fractured data.

The proposed quantities $\mu_{nm}$, $d_{nm}$, $C_{nm}^{[P]}$, and $\widehat C_{nm}^{[P]}$ form a morphometric layer between binary segmentation and physical modelling.

\section*{Statements and Declarations}

\textbf{Funding.} This research did not receive any specific grant from funding agencies in the public, commercial, or not-for-profit sectors.

\textbf{Competing interests.} No competing interests are declared.

\textbf{Data availability.} The primary micro-CT datasets used in this study are publicly available from the Imperial College Micro-CT Images and Networks collection and the Digital Porous Media Portal datasets DRP-317, DRP-5, DRP-31, and DRP-285, as cited in the reference list. Derived tables generated during the current study are available from the corresponding author upon reasonable request.

\textbf{Code availability.} The analysis scripts and derived tables are provided in the ancillary reproducibility archive submitted with this preprint.

\textbf{Use of generative AI tools.} During manuscript preparation, generative AI-assisted tools were used for language editing, code generation, and symbolic or numerical consistency checks. All generated material was reviewed and edited. The tools were not used as sources of primary scientific data or to create or alter figures or artwork.

\end{document}